# Semantic-based Detection of Segment Outliers and Unusual Events for Wireless Sensor Networks

(Research-in-Progress)

**Lianli Gao**

eResearch Lab, School of ITEE, The University of Queensland, Brisbane, Queensland 4072, Australia

l.gao1@uq.edu.au

**Michael Bruenig**

CSIRO Computational Informatics, Brisbane, Queensland 4069, Australia

Michael.Bruenig@csiro.au

**Jane Hunter**

eResearch Lab,School of ITEE, The University of Queensland, Brisbane, Queensland 4072, Australia

j.hunter@uq.edu.au

**Abstract:** Environmental scientists have increasingly been deploying wireless sensor networks to capture valuable data that measures and records precise information about our environment. One of the major challenges associated with wireless sensor networks is the quality of the data – and more specifically the detection of segment outliers and unusual events. Most previous research has focused on detecting outliers that are errors that are caused by unreliable sensors and sensor nodes. However, there is an urgent need for the development of new tools capable of identifying, tagging and visualizing erroneous segment outliers and unusual events from sensor data streams. In this paper, we present a SOUE-Detector (Segment Outlier and Unusual Event-Detector) system for wireless sensor networks that combines statistical analyses using Dynamic Time Warping (DTW) with domain expert knowledge (captured via an ontology and semantic inferencing rules). The resulting Web portal enables scientist to efficiently search across a collection of wireless sensor data streams and identify, retrieve and display segment outliers (both erroneous and genuine) within the data streams. In this paper, we firstly describe the detection algorithms, the implementation details and the functionality of the SOUE-Detector system. Secondly we evaluate our approach using data that comprises sensor observations collected from a sensor network deployed in the Springbrook National Park in Queensland, Australia. The experimental results show that the SOUE-Detector can efficiently detect segment outliers and unusual events with high levels of precision and recall.

**Key Words**: Dynamic Time Warping, Data Quality, Wireless Sensor Networks, Outlier Detection, Unusual Event Detection, Semantic Web.

## 1. INTRODUCTION

In recent years, wireless sensor networks (WSNs) have been increasingly deployed in environmental monitoring to measure physical and environmental parameters by densely deploying numerous tiny sensors in spatially distributed terrain (Corke et al., 2010, Selavo et al., 2007, Jelicic et al., 2011, Kułakowski et al., 2013, Sreedevi and Sebastian, 2012). Such WSNs have become a valuable tool as they can collect diverse observations ranging from air temperature and leaf wetness to images and acoustic telemetry data at high frequencies - providing detailed data which can be used by scientists, resource managers and policy makers (Wark et al., 2008, Jain et al., 2012). However, wireless sensor networks (WSN) comprising of hundreds of inexpensive and battery-operated sensors frequently suffer from poor data quality. The limited capabilities of low cost sensors (including battery power, memory, computational capacity, communication bandwidth) combined with harsh environmental conditions often generate unreliable and inaccurate data (Chen et al., 2006, Yang et al., 2010, Dereszynski and Dietterich, 2011, Janakiram et al., 2006). Noisy, faulty, missing and redundant data significantly affects the accuracy and reliability of the information inferred from WSN-based applications and its ability to be used in decision-making. Hence, it is essential to develop advanced algorithms and systems to precisely detect outliers within sensor data streams ("segments that significantly deviate from the normal pattern of sensed data") (Zhang, Meratnia & Havinga 2010) – and to distinguish between those outliers that are errors and those outliers that are "genuine" i.e., correctly measured data streams that represent unusual events.

Early research efforts aimed at detecting outliers have adopted: statistical-based approaches (Weili et al., 2007, Bettencourt et al., 2007); nearest neighbor-based techniques (Branch et al., 2006); clustering-based approaches (Rajasegarar et al., 2006); classification-based approaches (Rajasegarar et al., 2007, Hill et al., 2007, Janakiram et al., 2006) and Semantic Web-based approaches (Calder et al., 2010). However, these previous approaches have a number of shortcomings when applied to the detection of genuine outliers. Firstly, the majority of previous work assumes that the sensor data is univariate and fails to take into account multivariate data (Yang et al., 2010). Secondly, little effort has focused on handling the dynamic nature, variability and heterogeneity of sensors (the devices that detect or measure a physical property), sensor nodes (the platforms on which multiple sensors can be attached) and sensor networks. For example, in April 2008 the Springbrook Wireless Sensor Network (Wark et al., 2008) installed 9 sensor nodes with sensors measuring leaf wetness, soil moisture, air pressure, air temperature, relative humidity and wind (direction and speed). In April 2009, more sensor nodes were installed and new sensors including rainfall and light were added. In February 2011, an additional 125 sensor nodes were added and a range of new sensors started collecting information on tree growth, carbon dioxide concentrations, cloud cover and fog density. At the same time, the wind direction sensors were removed. Sensor nodes were moved between different locations at different points in time, and were re-configured with different numbers and types of sensors. Our approach will keep track of and update the configuration data of the WSN as it changes. Thirdly, using supervised machine learning algorithms such as Support Vector Machines (Zhang et al., 2013, Rajasegarar et al., 2007) and Bayesian Networks (Janakiram et al., 2006) to classify normal or abnormal data is difficult to achieve because there is no prior knowledge available via a training dataset. We will overcome this by designing new algorithms which take domain expert knowledge into account. Fourthly, correlations among sensor properties have been ignored by most of previous studies, but such correlations are very useful in detecting unusual events and can be captured through domain expert knowledge. Hence, our approach will include a tool to capture domain expert knowledge about correlations between sensor properties (e.g., between temperature, humidity, windspeed trends). Lastly and most importantly, most of the previous studies assume by default that any outliers are errors (Yang et al., 2010). They do not attempt to distinguish between the erroneous outliers and genuine outliers associated with unusual events (Shahid et al., 2012, Yang et al., 2010). Correctly distinguishing between outliers that are errors and outliers that are unusual events is critical for advancing the development and adoption of WSNs to underpin accurate and reliable decision support systems. In this study, an erroneous outlier, also called a true error, is defined as a segment of sensor data stream which refers to noise-related measurements or data generated by a faulty sensor. Such a segment deviates from the usual sensor data streams and is also likely to be spatially unrelated. An unusual event is defined as a particular phenomenon which simultaneously changes the patterns of multiple types of sensor data streams so that they deviate from the normal patterns of the data streams. In addition, the sensor data streams associated with unusual events are likely to be spatially or temporally correlated. Thus in this paper we focus on distinguishing between erroneous outliers and unusual events by making use of spatio-temporal and other types of correlations between sensor properties.

Semantic Web-based approaches, including RDF, ontologies and inferencing rules, have been applied previously to reason about anomalous sensor data (Calder et al., 2010). Specifically, Calder et al provided a framework for validating scientists' or decision makers' hypotheses about anomalous sensor data.

However, the identification of anomalous sensor data depends on scientists' or decision makers' prior definitions of what constitutes anomalous data (e.g. valid ranges). Compared with this work, we are not recording specific definitions about outliers or events from domain experts but documenting specific correlations between sensor properties to improve the detection of genuine outliers associated with unusual events. Moreover, we are not providing a platform for hypothesis validation but providing an effective approach to automatically detect outliers and unusual events for WSNs which monitor environmental variables. In addition, we provide intuitive user interfaces to enable domain scientists to express their knowledge about relationships or correlations between multiple sensor data streams measuring different properties. Our assumption is that unusual events can be identified by unusual patterns across multiple sensor properties if there is a correlation between those properties.

## 2. OBJECTIVES

The principal aim of this study is to develop a Segment Outlier and Unusual Event (SOUE) Detector to provide solutions to the issues outlined above. The more specific objectives are:

- To define a Correlation of Environmental Sensor Properties (CESP) ontology by extending the Semantic Sensor Network (SSN) ontology (Compton et al., 2012) developed by the Semantic Sensor Network Incubator Group to describe correlations between specific sensor properties. For example, *air temperature* has a correlation with *relative humidity* since air temperature increases as relative humidity decreases (Valsson and Bharat, 2011) .
- To design an algorithm to detect segment outliers and unusual events for WSNs by integrating statistical analysis techniques with Semantic Web technologies and domain expert knowledge about rules/correlations. This algorithm not only takes into account the dynamic nature and variability of sensors, sensor nodes and sensor networks, but also integrates information about other types of correlations apart from spatio-temporal correlations between sensor properties.
- To develop a SOUE-Detector system based on the designed algorithm that will validate our proposed methodology by applying it to a real dataset - Springbrook dataset (Wark et al., 2008). The system should enable users to search the Springbrook dataset and retrieve data streams for a particular time period and automatically tag segments as "error" or "unusual event" displayed in a visualization interface.

In the remainder of this paper, we describe the SOUE-Detector system in more detail. Section 3 describes the proposed methodology including the case study, data collection and our approach. Section 4 describes our proposed ontology of Correlated Environmental Sensor Properties (CESP). Section 5 provides the detailed information about our proposed detection algorithm for identifying erroneous outliers and unusual events. Section 6 describes the SOUE-Detector system architecture and Web portal. In Section 7, we evaluate the precision and recall of our proposed semantic-based algorithm. Finally, Section 8 provides a brief conclusion and future work.

## 3. CASE STUDY

### *3.1 Case Study and Data Collection*

The Springbrook Wireless Sensor Network (Wark et al., 2008) project (implemented by CSIRO, Queensland Department of Environment and Resource Management and the Australia Rainforest Conservation Society) involves the deployment of a WSN in the Springbrook National Park, located about 96km south of Brisbane in the state of Queensland, Australia. It aims to provide a research platform for monitoring how the microclimates and biodiversity of the Springbrook plateau changes over time. Hundreds of solar-powered sensor nodes have been installed and each sensor node carries several sensing devices to collect different environmental variables. These variables include air temperature, relative humidity, air pressure, leaf wetness, soil moisture, wind speed, wind direction and light. The richness of the Springbrook data set makes it ideal as a testbed for developing and evaluating our approach of detecting outliers and unusual events. For our case study, we acquired 2.5 years' data (from January 2010 to June 2012) from the Springbrook Wireless Sensor Network project. More specifically, three types of sensor observations (air temperature, relative humidity and air pressure) were collected by Vaisla WXT520 weather transmitter. The sampling rate of each sensor was 1 sample/10 min.

### *3.2 Approach*

Our approach to developing and applying our system to the case study data can be sub-divided into the following five steps:

1. Development of a PostgreSQL database to store raw sensor data streams collected from the WSNs. Specifically, each sensor observation includes the following metadata: the ID of the sensor node, the ID of the sensor device, the name of the sensor device, the sampling time/date for the observation, the type of sensor property, the observation value and the observation measurement unit. Each sensor node has the following metadata: the ID of the sensor node, its geographical location (latitude and longitude), its installation date/time and removal date/time. In addition, each sensor device has the following metadata: the ID of the sensor, the ID of sensor node that the sensor is connected to, sensor observation property, installation date/time and removal date/time.
2. Development of an ontology of Correlated Environmental Sensor Properties (CESP). This ontology defines concepts that describe different correlation types in terms of strength, direction, shape, space-time, composition and complexity.
3. Design and implementation of the detection algorithm, which comprises: definition of a spatial neighborhood matrix for describing the spatial correlations between sensor nodes; definition of spatial neighborhood matrices for describing the spatial correlations between sensors; detection of suspicious data streams (using Dynamic Time Warping-based similarity computation); development of a user interface for domain experts to define correlation rules; identification of unusual events from suspicious data streams.
4. Development of a Web-based system to enable users to search wireless sensor data streams for a given time period, then to visualize the detected erroneous outliers and unusual events via a Google Earth Map and timeline interface.
5. Evaluation of the performance of the proposed detection algorithm by calculating the precision and recall on two datasets.

## 4. AN ONTOLOGY OF CORRELATED ENVIRONMENTAL SENSOR PROPERTIES

Most of the sensor properties have one or more correlations with other properties. For example, humidity and barometric pressure are related to air temperature. Capturing the correlations between sensor properties is critical for improving the accuracy and efficiency of detecting genuine outliers and unusual events. Hence, the first step in detecting outliers and events is to develop a Correlated Environmental Sensor Properties (CESP) ontology (https://code.google.com/p/cesp-ontology/source/browse/wiki/cesp-v1.owl) that describes the correlations between environmental sensor properties. The CESP is designed by extending the Semantic Sensor Network Ontology (SSN) that has been developed by the W3C Semantic Sensor Network Incubator Group (Compton et al., 2012) for describing sensors, sensing, the measurement capabilities of sensors and the observations. More specifically, the SSN ontology is a high level ontology that does not describe the low-level properties of a wireless sensor network, such as the relationships between sensor properties. Hence for the work presented in this paper, we have developed the CESP ontology by extending the SSN ontology to precisely describe the specific sensor properties and their relationships within a sensor network.

The CESP ontology (See in Fig.1) consists of two components: a sensor property class (*ssn:Property*) and an object property (*cesp:Correlation*) which is defined as a relationship between two sensor properties. The subclasses of *ssn:Property* are derived from the Climate and Forecast (CF) metadata conventions (Eaton et al., 2003) that provide a definitive description of climate and forecast data variables and the spatial and temporal properties of the variables. For example, for each of the CF names (e.g., *air_temperature*) we created a corresponding class e.g., *cesp:air_temperature*. A sub-class of *ssn:Property* (e.g. *cesp:air_temperature, cesp:relative_humidity, cesp:air_pressure*) must be observed by a sensor. The property *cesp:Correlation* links one sensor property to another sensor property. In order to accurately describe the correlations between sensor properties, we defined five types of sub-properties of *cesp:Correlation* (Tutorvista, 2013, Zone, 2013b, Zone, 2013a):

1. Strength – *cesp:hasVeryStrongCorrelation, cesp:hasStrongCorrelation, cesp:hasMediumCorrrelation, cesp:hasWeakCorrelation, cesp:hasVeryWeakCorrelation, cesp:hasVeryWeakCorrelation*
2. Direction – *cesp:hasPositiveCorrelation, cesp:hasNegativeCorrelation*
3. Shape/Form – *cesp:hasLinearCorrelation, cesp:hasCurvilinearCorrelation, cesp:hasScatteredCorrelation*
4. Space-time – *cesp:hasSpatialCorrelation, cesp:hasTemporalCorrelation, cesp:hasSpatioTemporalCorrelation*
5. Composition – *cesp:hasPartialCorrelation, cesp:hasSimpleCorrelation, cesp:hasMultipleCorrelation*

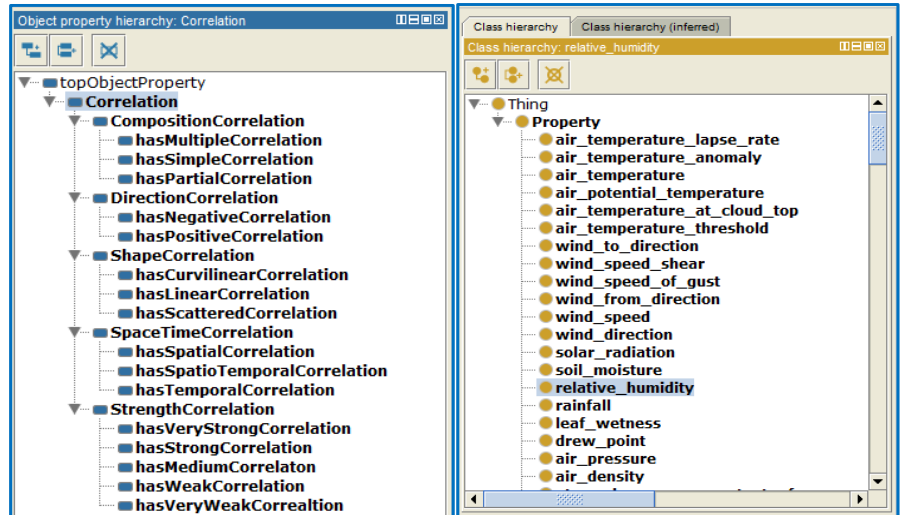

Fig.1: The CESP ontology consists of an object property *cesp:Correlation* with a set of sub object properties (left) and a sensor property class *ssn:Property* with a set of sub classes (right)

## 5. DETECTION ALGORITHMS

In this section, we describe the algorithms that used for detecting outliers and unusual events. To begin, we define the terms and notations used in our algorithms, and also describe the similarity computation method which uses Dynamic Time Warping (DTW) to compute the similarity between two windows of data streams. Following this, we describe the five steps in our detection algorithm:

1. We construct a sensor node matrix to document the spatial neighborhood relationships between sensor nodes by defining a threshold value that determines whether two sensor nodes are neighbors or not. If the distance between two sensor nodes is <= the threshold value, then they are neighbors. If the distance between two sensor nodes is > than the threshold value, then they are not neighbors.
2. We construct a matrix for the sensors to document spatial relationships between sensors. Such matrices are built by integrating the sensor node matrix constructed in the first step with sensor node configurations (that define which sensors are attached to which sensor nodes) of a WSN.
3. Next, we detect suspicious data within the sensor data streams by: choosing a sensor type (e.g. temperature) and determining the similarities between neighbouring data streams for each sensor type. The algorithm determines whether the sensor collects suspicious data or not by applying a predefined rule to the calculated similarity results. This is repeated for all sensors in the network.
4. We enable domain experts to input/define correlation rules through a graphical user interface. The user interface enables a user to specify relationships (e.g. very strong, strong, medium, weak, very weak) among sensor properties. The specified relationships are saved and used in the next step.
5. Finally, we combine the domain expert rules (from the previous step) with the detected suspicious results to determine whether a suspicious segment data stream is a true error or an unusual event.

## *5.1 Notations and Definitions*

Let $\{s_i\}_{i=1}^m = \{s_1, s_2, \cdots, s_m\}$ denote all the sensors deployed in the WSN, where $s_i = (s_{i1}, s_{i2}, \cdots, s_{in})$ indicates the i-th type of sensor employed in the WSN, $s_{ij}$ indicates the i-th type of sensor installed on the j-th sensor node $sn_j$, and n is the total number of sensor nodes deployed in the WSN. For instance, $s_1$ denotes all the relative humidity sensors deployed in the WSN and $s_2$ denotes all the air temperature sensors deployed in the WSN. During a time period $t$, $s_{ij}$ has collected a set of sensor observations $\mathbf{O}_t^{ij} = \begin{pmatrix} t_1, \cdots, t_g \\ o_1^{ij}, \cdots, o_g^{ij} \end{pmatrix}^T$ where g is the total number of collected sensor observations. The complete set of sensor observations collected by the $s_i$ during the time $t$ are expressed as $\mathbf{O}_t^i = (\mathbf{O}_t^{i1}, \cdots, \mathbf{O}_t^{ij}, \cdots, \mathbf{O}_t^{in})$.

## *5.2 Dynamic Time Warping based Similarity Computation*

We use Dynamic Time Warping (DTW) (Sakoe and Chiba, 1990, Keogh, 2002, Matuschek et al., 2008) to compute the similarity between two sensor observation data streams, as it is a well-established and widely used algorithm for comparing similarity between two discrete sequences of continuous values. Suppose that two sensors $s_{ij}$ and $s_{ik}$ have been installed in the WSN, and $s_{ik} \in NH(s_{ij})$ and $s_{ij} \in NH(s_{ik})$, where $NH(sn_i)$ indicates the neighbour sensor nodes of $sn_i$. During a short time period $\Delta t$, $s_{ij}$ and $s_{ik}$ both respectively collected u sensor observations: $\mathbf{O}_{\Delta t}^{ij} = \begin{pmatrix} t_1, \cdots, t_u \\ o_1^{ij}, \cdots, o_u^{ij} \end{pmatrix}^T$ and $\mathbf{O}_{\Delta t}^{ik} = \begin{pmatrix} t_1, \cdots, t_u \\ o_1^{ik}, \cdots, o_u^{ik} \end{pmatrix}^T$, where $t_1, t_2, \cdots, t_u$ ($t_1 \prec t_2 \prec \cdots \prec t_u$) are the timestamps at when sensor observations are collected, and $o_1^{ij}, o_2^{ij}, \cdots, o_u^{ij}$ and $o_1^{ik}, o_2^{ik}, \cdots, o_u^{ik}$ are the corresponding captured sensor observations. In order to efficiently compute the trend similarity $sim(\mathbf{O}_{\Delta t}^{ij}, \mathbf{O}_{\Delta t}^{ik})$ between $\mathbf{O}_{\Delta t}^{ij}$ and $\mathbf{O}_{\Delta t}^{ik}$, we project them to a two dimensional Cartesian coordinate system where we treat time $t_\lambda$ ($1 \leq \lambda \leq u$) as x-coordinate value and observations $o_\lambda^{ij}$ ($o_\lambda^{ik}$) as y-coordinate value. The difference $(t_{\lambda+1} - t_\lambda, o_{\lambda+1}^{ij} - o_\lambda^{ij})$ ($(t_{\lambda+1} - t_\lambda, o_{\lambda+1}^{ik} - o_\lambda^{ik})$) between two connecting successive point can be expressed as a vector $\mathbf{v}_\lambda^{ij} = (t_{\lambda+1} - t_\lambda, o_{\lambda+1}^{ij} - o_\lambda^{ij})$ ($\mathbf{v}_\lambda^{ik} = (t_{\lambda+1} - t_\lambda, o_{\lambda+1}^{ik} - o_\lambda^{ik})$). Eventually, these two sensor observation data sets are respectively transformed to two sets of vector sequences: $\mathbf{v}^{ij} = (\mathbf{v}_1^{ij}, \cdots, \mathbf{v}_a^{ij}, \cdots, \mathbf{v}_{u-1}^{ij})$ and $\mathbf{v}^{ik} = (\mathbf{v}_1^{ik}, \cdots, \mathbf{v}_b^{ik}, \cdots, \mathbf{v}_{u-1}^{ik})$, where $1 \leq a \leq (u-1)$ and $1 \leq b \leq (u-1)$.

Using the DTW, a $(u-1)\times(u-1)$ matrix can be obtained, where element $(a,b)$ can be computed by Euclidean Distance between the end points of the two vectors or the angle $\theta$ ( $0\leq\theta\leq\pi$) between two vectors (Toshniwal and Joshi, 2005). However, the Euclidean Distance is not able to handle vertical shift existing between the vectors under comparison. Compared with the Euclidean Distance, the angle not only considers the direction of the vectors, but also handles the vertical shift between the vectors. Thus, in this work we adopt the angle $\theta$ between two vectors (e.g. $\mathbf{v}_a^{ij}$ and $\mathbf{v}_b^{ik}$) to compute the degree of similarity. More specifically, the angle $\theta\left(\mathbf{v}_a^{ij}, \mathbf{v}_b^{ik}\right)$ between vector $\mathbf{v}_a^{ij}$ and $\mathbf{v}_b^{ik}$ is defined as:

$$\theta\left(\mathbf{v}_a^{ij}, \mathbf{v}_b^{ik}\right) = \mathrm{acos}\frac{\mathbf{v}_a^{ij}\bullet\mathbf{v}_b^{ik}}{\left|\mathbf{v}_a^{ij}\right|\left|\mathbf{v}_b^{ik}\right|} \quad (1)$$

Then, an alignment between $\mathbf{O}_{\Delta t}^{ij}$ and $\mathbf{O}_{\Delta t}^{ik}$ can be represented by a warping path $W = \{w_1, w_2, \cdots, w_\ell, \cdots, w_K\}$, where $1\leq\ell\leq K$ and $(u-1)\leq K\leq(2u-3)$. For each $w_\ell$, it must satisfy three constraints: Boundary condition, Continuity condition and Monotonic condition (Sakoe and Chiba, 1990, Al-Naymat et al., 2009, Ongwattanakul and Srisai, 2009). In fact, many possible monotonical alignment paths from $(1,1)$ to $(u-1,u-1)$ can be generated to meet the three constraints. To determine an optimal warping path to minimize the cumulated distance $D(a,b)$ between $\mathbf{O}_{\Delta t}^{ij}$ and $\mathbf{O}_{\Delta t}^{ik}$, a dynamic programming algorithm is an effective approach:

$$D(1,1) = 0 \quad (2)$$

$$D(a,b) = \min\{D(a\text{-}1,b\text{-}1), D(a\text{-}1,b), D(a,b\text{-}1)\} + \theta\left(\mathbf{v}_a^{ij}, \mathbf{v}_b^{ik}\right) \quad (3)$$

With $D(a,b)$, we use the following equation to calculate the similarity between $\mathbf{O}_{\Delta t}^{ij}$ and $\mathbf{O}_{\Delta t}^{ik}$ is:

$$\mathrm{sim}\left(\mathbf{O}_{\Delta t}^{ij}, \mathbf{O}_{\Delta t}^{ik}\right) = \begin{cases} 0, & \text{if } \dfrac{D(u-1,u-1)}{K} \succ \dfrac{\pi}{2} \\ \cos\left(\dfrac{D(u-1,u-1)}{K}\right), & \text{otherwise} \end{cases} \quad (4)$$

## 5.3 Construction of Spatial Neighbourhood Matrix for Sensor Nodes

In this section, we describe our approach for constructing a matrix $\mathbf{U}$ that documents the spatial neighbourhood relationship among sensor nodes deployed in the WSN. Formally, $\mathbf{U}$ can be expressed as:

$$\mathbf{U} = \begin{pmatrix} u_{11} & \cdots & u_{1n} \\ \vdots & \ddots & \vdots \\ u_{n1} & \cdots & u_{nn} \end{pmatrix} \in \mathbb{R}^{n\times n} \quad (5)$$

where $u_{ij}\in\{0,1\}$ ($0\leq i\leq n$ and $0\leq j\leq n$). If $u_{ij}=1$, it indicates that $sn_j \in NH(sn_i)$. If $u_{ij}=0$, it indicates that $sn_j \notin NH(sn_i)$. The distance between $sn_i$ and $sn_j$ (each located at a precise latitude and longitude: $(lat_i, long_i)$ and $(lat_j, long_j)$) is $D(sn_i, sn_j)$ and $\partial$ is a predefined neighbourhood threshold value. To build $\mathbf{U}$, if $D(sn_i, sn_j)\leq\partial$, then $u_{ij}=1$. If $D(sn_i, sn_j)\succ\partial$, then $u_{ij}=0$. Specifically,

$$D(sn_i, sn_j) = 2rb \quad (6)$$

where $r = 6,378,137$ is the diameter of the earth in meters according to WSG84 system and $b$ is defined as:

$$b = a\tan 2\left(\sqrt{d}, \sqrt{1-d}\right) \quad (7)$$

where $d$ is defined as:

$$d = \sin\left(\frac{\pi}{360}(\text{lat}_i - \text{lat}_j)\right) * \sin\left(\frac{\pi}{360}(\text{lat}_i - \text{lat}_j)\right) + \cos\left(\frac{\pi}{180}\text{lat}_i\right)\cos\left(\frac{\pi}{180}\text{lat}_j\right)$$
$$* \sin\left(\frac{\pi}{360}(\text{long}_i - \text{long}_j)\right) * \sin\left(\frac{\pi}{360}(\text{long}_i - \text{long}_j)\right) \quad (8)$$

## 5.4 Construction of Spatial Neighbourhood Matrices for Sensors

In this section, we detail our algorithm for constructing spatial neighbourhood matrices for sensors. For i-th type of sensors $\mathbf{s}_i = (s_{i1}, s_{i2}, \cdots, s_{in})$ where $1 \leq i \leq m$, a corresponding spatial neighbourhood matrix $\mathbf{A}_i \in \mathbb{R}^{n \times n}$ will be constructed. There are m types of sensors in total, thus m matrices $\{\mathbf{A}_i\}_{i=1}^{m}$ will be constructed in total. Formally,

$$\mathbf{A}_i = \mathbf{E}_i \bullet \mathbf{U} = \begin{pmatrix} (a_i)_{11} & \cdots & (a_i)_{1n} \\ \vdots & \ddots & \vdots \\ (a_i)_{n1} & \cdots & (a_i)_{nn} \end{pmatrix} \in \mathbb{R}^{n \times n} \quad (9)$$

where $(a_i)_{jk} \in \{0,1\}$ ( $1 \leq k \leq n$ ) describes the spatial neighbourhood relationship between two sensors $s_{ij}$ and $s_{ik}$. If $s_{ik}$ and $s_{ij}$ exist, and $s_{ik} \in \text{NH}(s_{ij})$, then $(a_i)_{jk} = 1$. If either $s_{ik}$ or $s_{ij}$ does not exist, or $s_{ik} \notin \text{NH}(s_{ij})$, then $(a_i)_{jk} = 0$. In addition,

$$\mathbf{E}_i = (\mathbf{e}_i)^T \mathbf{e}_i = \begin{pmatrix} (e_i)_1 * (e_i)_1 & \cdots & (e_i)_1 * (e_i)_n \\ \vdots & \ddots & \vdots \\ (e_i)_n * (e_i)_1 & \cdots & (e_i)_n * (e_i)_n \end{pmatrix} = \begin{pmatrix} (e_i^{'})_{11} & \cdots & (e_i^{'})_{1n} \\ \vdots & \ddots & \vdots \\ (e_i^{'})_{n1} & \cdots & (e_i^{'})_{nn} \end{pmatrix} \in \mathbb{R}^{n \times n} \quad (10)$$

where

$$\mathbf{e}_i = \left((e_i)_1, \cdots, (e_i)_j, \cdots, (e_i)_n\right) \in \mathbb{R}^{1 \times n} \quad (11)$$

where $(e_i)_j \in \{0,1\}$ and $1 \leq j \leq n$. If $(e_i)_j = 1$, it indicates that $s_{ij}$ is employed in the WSN. If $(e_i)_j = 0$, it indicates that sensor node $sn_j$ did not have an i-th type of sensor installed on.

## 5.5 DTW-based Similarity Matrices Construction and Suspicious Data Detection

Given that m types of sensors deployed in the WSN have collected a set of sensor observations $\{\mathbf{O}_t^i\}_{i=1}^{m} = \{\mathbf{O}_t^{i1}, \cdots, \mathbf{O}_t^{ij}, \cdots, \mathbf{O}_t^{in}\}_{i=1}^{m}$ during a time period t. If sensor $s_{ij}$ exists, then $\mathbf{O}_t^{ij} = \begin{pmatrix} t_1, \cdots, t_g \\ o_1^{ij}, \cdots, o_g^{ij} \end{pmatrix}^T$. If $s_{ij}$ does not exist, then $\mathbf{O}_t^{ij} = []$. To detect suspicious data, firstly we need to construct DTW-based similarity matrices $\{\mathbf{SM}_i\}_{i=1}^{m}$ for the WSN, where $\mathbf{SM}_i$ is a matrix describing all the similarity relationships between two elements of $\{\mathbf{O}_t^{i1}, \cdots, \mathbf{O}_t^{ij}, \cdots, \mathbf{O}_t^{in}\}$:

$$\mathbf{SM}_i = \begin{pmatrix} (sm_i)_{11} & \cdots & (sm_i)_{1n} \\ \vdots & \ddots & \vdots \\ (sm_i)_{n1} & \cdots & (sm_i)_{nn} \end{pmatrix} \quad (12)$$

where $(sm_i)_{jk}$ ( $1 \leq j \leq n$ and $1 \leq k \leq n$ ) describes the similarity trend between $\mathbf{O}_t^{ij}$ and $\mathbf{O}_t^{ik}$. Specifically, if $(a_i)_{jk} = 0$, then $(sm_i)_{jk} = []$. If $(a_i)_{jk} = 1$, then the $(sm_i)_{jk}$ is represented as:

$$(\mathbf{sm}_i)_{jk} = \left(\left((sm_i)_{jk}\right)_1, \cdots, \left((sm_i)_{jk}\right)_\ell, \cdots, \left((sm_i)_{jk}\right)_\kappa\right) \in \mathbb{R}^{1\times\kappa} \quad (13)$$

where $1 \leq \ell \leq \kappa$. To compute the $(\mathbf{sm}_i)_{jk}$, we take two sensor observations data sets $\mathbf{O}_t^{ij}$ and $\mathbf{O}_t^{ik}$ as input, then divide them into $\kappa$ sliding windows, where $\kappa = \frac{2g}{\eta} - 1$ ( g is the number of sensor observations and $\eta$ is an even integer denotes the predefined size of a sliding window). In addition, each window contains an overlap of $\frac{\eta}{2}$ observations between consecutive windows. Finally, the $\mathbf{O}_t^{ij}$ and $\mathbf{O}_t^{ik}$ are converted to:

$$\mathbf{O}_t^{ij} = \left\{\mathbf{O}_{\Delta t_1}^{ij}, \cdots, \mathbf{O}_{\Delta t_\ell}^{ij}, \cdots, \mathbf{O}_{\Delta t_\kappa}^{ij}\right\} \quad (14)$$

$$\mathbf{O}_t^{ij} = \left\{\mathbf{O}_{\Delta t_1}^{ik}, \cdots, \mathbf{O}_{\Delta t_\ell}^{ik}, \cdots, \mathbf{O}_{\Delta t_\kappa}^{ik}\right\} \quad (15)$$

where $\Delta t_\ell$ indicates a short time period $[t_{(\ell-1)*\eta+1}, t_{\ell*\eta+1}]$, $\mathbf{O}_{\Delta t_\ell}^{ij} = \begin{pmatrix} t_{(\ell-1)*\eta+1}, \cdots, t_{\ell*\eta+1} \\ o_{t_{(\ell-1)*\eta+1}}^{ij}, \cdots, o_{t_{\ell*\eta+1}}^{ij} \end{pmatrix}^T$ and $\mathbf{O}_{\Delta t_\ell}^{ik} = \begin{pmatrix} t_{(\ell-1)*\eta+1}, \cdots, t_{\ell*\eta+1} \\ o_{t_{(\ell-1)*\eta+1}}^{ik}, \cdots, o_{t_{\ell*\eta+1}}^{ik} \end{pmatrix}^T$. The $\left((sm_i)_{jk}\right)_\ell$ is computed by Eq.(4), where $\left((sm_i)_{jk}\right)_\ell = sim(\mathbf{O}_{\Delta t_\ell}^{ij}, \mathbf{O}_{\Delta t_\ell}^{ik})$.

Once $\{\mathbf{SM}_i\}_{i=1}^m$ is obtained, it can be readily used to detect suspicious data. To detect suspicious data, firstly we need to construct a set of suspicious matrices $\mathbf{P} = \{\mathbf{P}_1, \cdots, \mathbf{P}_m\}$, where $\mathbf{P}_i = \left((\mathbf{p}_i)_1, \cdots, (\mathbf{p}_i)_j, \cdots, (\mathbf{p}_i)_n\right)$ ($1 \leq j \leq n$). Specifically, if $s_{ij}$ does not exist, then $(\mathbf{p}_i)_j = []$. If $s_{ij}$ exist, then

$$(\mathbf{p}_i)_j = \left\{(p_i)_{j1}, \cdots, (p_i)_{j\ell}, \cdots, (p_i)_{j\kappa}\right\} \in \mathbb{R}^{1\times\kappa} \quad (16)$$

where $(p_i)_{j\ell} \in \{1, 0\}$ and $1 \leq \ell \leq \kappa$. If $(p_i)_{j\ell} = 0$, it indicates that the segment of data streams $\mathbf{O}_{\Delta t_\ell}^{ij}$ is normal. If $(p_i)_{j\ell} = 1$, it indicates that the $\mathbf{O}_{\Delta t_\ell}^{ij}$ is suspicious. To compute the $(p_i)_{j\ell}$, we use the following function:

$$(p_i)_{j\ell} = \begin{cases} 0 & \text{if } \sum_{k=0}^n Z\left(\left((sm_i)_{jk}\right)_\ell\right) \geq \frac{\sum_{k=0}^n S\left(\left((sm_i)_{jk}\right)_\ell\right)}{2} \\ 1 & \text{otherwise} \end{cases} \quad (17)$$

Where Z and S are:

$$Z(x) = \begin{cases} 0, & x \prec \beta \\ 1, & \text{otherwise} \end{cases} \quad (18)$$

$$S(x) = \begin{cases} 0, & \text{if } x == 0 \\ 1, & \text{otherwise} \end{cases} \quad (19)$$

where $\beta$ is the predefined similarity threshold value. The rationale for choosing the function in Equation 17 is that: if a data stream segment of sensor (property A) exists, and its pattern is similar to less than half of its neighbors' patterns (for the same property A), then it is suspicious, otherwise it is normal.

## *5.6 Capturing Domain Expert Knowledge through Correlation Definition*
To capture domain experts' knowledge about trends between properties, we provide them with the Protégé ontology editing tool and user interface to define sensor property correlations using terms from the CESP ontology. Fig.2 shows a screen shot of an expert using Protégé to create a correlation (*cesp:air_temperature cesp:hasNegativeCorrealtion cesp:relative_humidity*) which specifies that air temperature has a negative (inverse) correlation on relative humidity. Fig.2 also shows that air temperature has a strong correlation with relative humidity (*cesp:air_temperature cesp:hasStrongCorrelation cesp:relative_humidity*). After the domain experts have defined the correlations between properties, we store this information in the sensor property knowledge base (an RDF triple store).

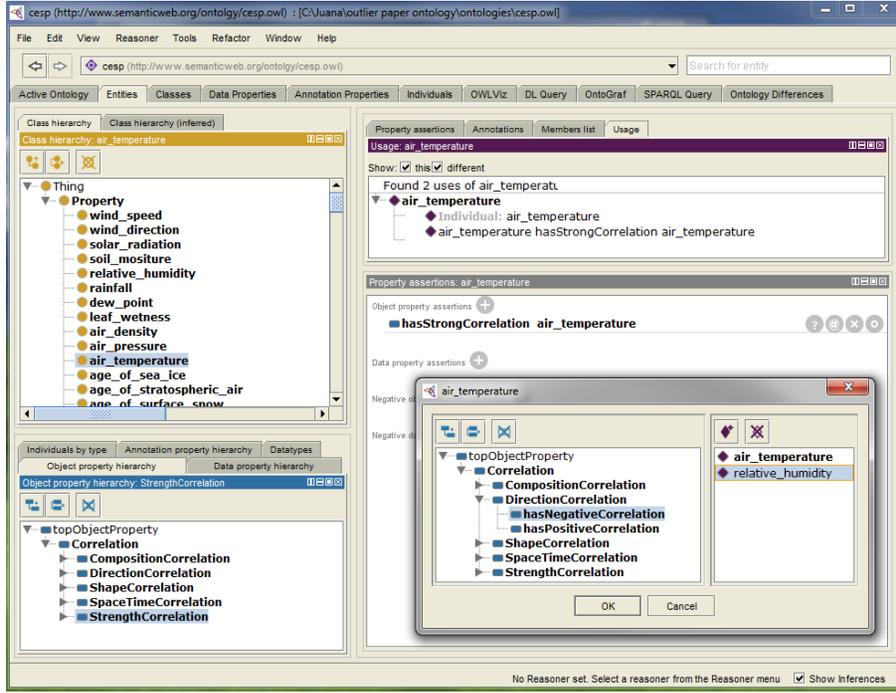

Fig.2: An example of an expert defining a strong *cesp:hasNegativeCorrelation*

Given these domain expert rules, we can then construct a relationship matrix $\mathbf{Y}$. Formally, $\mathbf{Y}$ is defined as

$$\mathbf{Y} = \begin{pmatrix} y_{11} & \cdots & y_{1m} \\ \vdots & \ddots & \vdots \\ y_{m1} & \cdots & y_{mm} \end{pmatrix} \in \mathbb{R}^{m \times m} \quad (20)$$

where $y_{ii'} \in \{0,1\}$ describes the correlation between i-th sensor property and i'-th sensor property, $1 \leq i \leq m$ and $1 \leq i' \leq m$. If i-th sensor property has a specific correlation (e.g. strong correlation or medium correlation (defined by an expert)) with the i'-th sensor property, then $y_{ii'} = 1$, otherwise $y_{ii'} = 0$. In addition, if $i' = i$, then $y_{ii'} = y_{i'i} = 0$. The value of $y_{ii'}$ is obtained by submitting a SPARQL query to the sensor property knowledge base. Take the strong and medium correlation as an example, if i-th sensor property stands for the air temperature property, and i'-th sensor property stands for the relative humidity property, then we submit the following SPARQL query to the sensor property knowledge base to calculate the value of $y_{ii'}$:

ASK { {cesp:air_temperature cesp:hasStrongCorrelation cesp:relative_humidity. } UNION {cesp:air_temperature cesp:hasMediumCorrelation cesp:relative_humidity.}}

If the query returns true, then $y_{ii'} = 1$. If the query returns false, then $y_{ii'} = 0$. Once $\mathbf{Y}$ is fully calculated, it can then be used to detect segment outliers and unusual events.

### *5.7 Detection of Segment Outliers and Unusual Events*
Finally we are able to define and apply rules to distinguish between segment outliers that are errors and genuine outliers (unusual events). For example: If a segment of observations of sensor property A is identified as suspicious data, and >50% of the corresponding segments of sensor properties that have a medium or strong correlation with A are also identified as suspicious data, then we infer that an unusual event has occurred. Otherwise, this segment of observations is an erroneous outlier. Specifically, for each $(p_i)_{j\ell}$ of $\mathbf{P}_i = \{(\mathbf{p}_i)_1, \cdots, (\mathbf{p}_i)_j, \cdots, (\mathbf{p}_i)_n\}$ of $\{\mathbf{P}_i\}_{i=1}^m$, if $(\mathbf{p}_i)_j \neq []$, then we need to determine whether the segment of observation $\mathbf{O}_{\Delta t_\ell}^{ij}$ contains any outliers and unusual events. If $(p_i)_{j\ell} = 0$, $\mathbf{O}_{\Delta t_\ell}^{ij}$ is a segment of normal data. If $(p_i)_{j\ell} = 1$, then we use the following function to determine whether $\mathbf{O}_{\Delta t_\ell}^{ij}$ indicates an erroneous outlier or an unusual event:

$$r_\ell^{ij} = f\left(\mathbf{O}_{\Delta t_\ell}^{ij}\right) = \begin{cases} R_1, & \text{if } C_1 \geq \dfrac{C_2}{2} \quad (R_1 \text{ indicates an unusual event}) \\ R_2, & \text{otherwise} \quad (R_2 \text{ indicates a erroneous outlier}) \end{cases} \quad (21)$$

where

$$C_1 = \sum_{i'=1}^m F\left((p_{i'})_{j\ell}, y_{ii'}\right) \quad (22)$$

$$C_2 = \sum_{i'=1}^m F_R\left((p_{i'})_{j\ell}, y_{ii'}\right) \quad (23)$$

where

$$F(x, y) = \begin{cases} 1, & \text{if } x==1 \text{ and } y==1 \\ 0, & \text{otherwise} \end{cases} \quad (24)$$

and

$$F_R(x, y) = \begin{cases} 1, & \text{if } x \text{ is not null}, y==1 \\ 0, & \text{otherwise} \end{cases} \quad (25)$$

Therefore, for each observation $\mathbf{O}_t^{ij}$ a decision matrix $\mathbf{r}_t^{ij} = \left(r_1^{ij}, \cdots, r_\ell^{ij}, \cdots r_\kappa^{ij}\right)$ will be constructed. If $r_\ell^{ij} = R_1$ then $\mathbf{O}_{\Delta t_\ell}^{ij}$ indicates an unusual events. If $r_\ell^{ij} = R_2$, then $\mathbf{O}_{\Delta t_\ell}^{ij}$ is an erroneous segment outlier. Finally, a decision matrix $\{\mathbf{R}_i\}_{i=1}^m$ with $\mathbf{R}_i = \left(\mathbf{r}_t^{i1}, \cdots, \mathbf{r}_t^{in}\right)$ is constructed. In addition, the computational complexity for detecting segment outliers and unusual events is $O(m*n^2)$.

# 6. IMPLEMENTATION

## 6.1 System Architecture
Fig.3 provides an overview of the architecture of the SOUE-Detector system. The system utilizes the PostgreSQL object-relational database management system for storing sensor observations and the open source Java framework Sesame, an RDF triple repository, for storing sensor property knowledge base. The SOUE-Detector Web portal provides a Web interface to enable users to search for sensor observations for a particular time period, and view the corresponding segment outliers and unusual events. The server component is built using JSP and Java. The server interfaces with users through a Web browser (e.g. Google Chrome) and a Google Earth map interface and a Google line chart interface, enabling spatial-temporal search and visualization across the data.

In addition, the PostgreSQL database stores the WSN configurations and associated sensor node matrix and sensor matrices, which document the spatial neighborhood relationship among sensor nodes and sensors, respectively. Date/time stamps are also recorded with these matrices, which are updated/recalculated whenever the WSN configuration is changed. Whenever the system receives a sensor network modification message, the system attaches end date/time stamps to the previously generated sensor node and sensor matrices and then recalculates the new sensor node and sensor matrices and saves them in the database with the new current start date/time stamps. Past matrices are not deleted because they are relevant for processing of historical archived data streams.

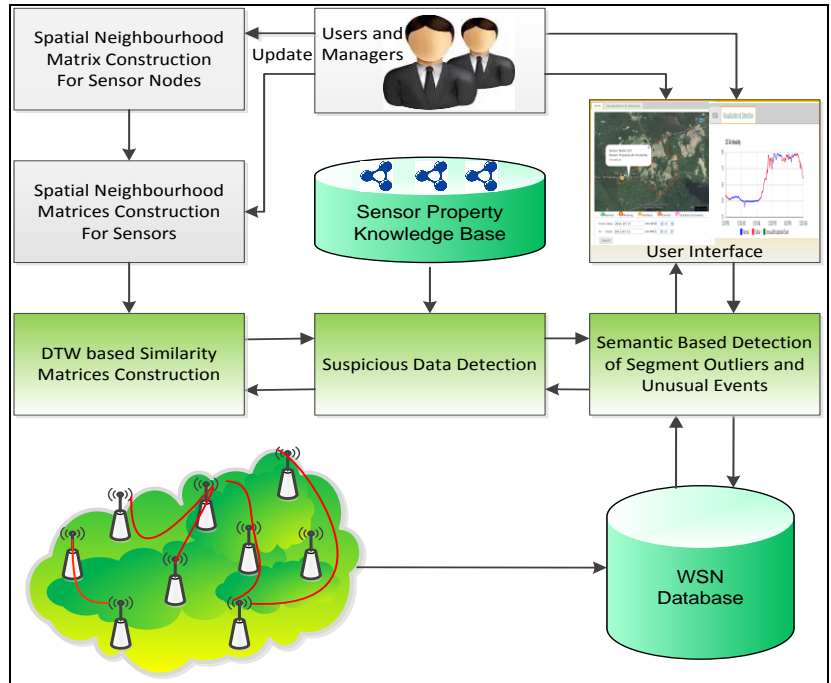

Fig.3: High level architecture view of the SOUE-Detector system

## *6.2 Web Portal and User Interface*

A Web Portal was developed to provide access to the wireless sensor database, RDF triple store and associated services. Fig.4 illustrates the screenshots of the search interface, the Google Earth interface and the timeline interface. Users are able to specify a time period of interest. The datasets within this time range are retrieved and our detection algorithm is applied to the search results to detect segment outliers. The detection results are displayed in a visualization/timeline interface. The Google Earth map visualization interface uses different sensor icons to represent each sensor's status. For example, a blue icon indicates a sensor with normal data, a yellow icon indicates a sensor with segment outliers, and an orange icon indicates a sensor with unusual events. The timeline visualization interface uses different colors to mark normal data (grey color), erroneous segment outliers (red color) and unusual events (royal blue).

The LHS of Fig.4 illustrates an example with erroneous segment outliers. A user specified a time period between 2011-06-04 00:00:00 and 2011-06-04 16:00:00. The system retrieved the data streams for this period for all sensors/sensor nodes, and then applied our detection algorithm to the search results. The system detects outliers in the data from Sensor node 1 (a relative humidity and an air temperature sensor). The detection results are displayed in a Google Map interface in Fig.4 (a-1). Fig.4 (a-2) shows the sensor data collected from *sensor 1-relative humidity*, displayed within a timeline with segment outliers marked as red. Clicking the "*Show Neighborhood Sensors*" button, retrieves data from the neighborhood sensors, including relative humidity sensors 26, 143 and 210, and displays these data streams as well (See Fig.4 (a-2)). Fig.4 (a-3) shows the detection results for *sensor 1-air temperature* and its neighborhood sensors collections. We know that the relative humidity property is strongly correlated with the air temperature. By comparing Fig.4 (a-2)) and Fig.4 (a-3), the user can see that:

- Relative humidity sensor 1 is different from the similar pattern displayed by its neighborhood sensors of relative humidity sensors (26, 143, 210) – so it is identified as a true error;

- The pattern of air temperature sensor 1 is similar to its neighborhood sensors' patterns (air temperature sensors 143, 210 and 26) – so it is normal data.

The RHS of Fig.4 illustrates an example of an unusual event detection. A user specified a time period between 2011-06-25 00:00:00 and 2011-06-25 10:00:00. The detection results are represented in Fig.4 (b-1), (b-2) and (b-3). From Fig.4 (b-1), we can see that some unusual event happened at sensor node 10. Selecting relative humidity sensor 10 and air temperature sensor 10 (and their neighborhood sensors), generates the results displayed in Fig.4 (b-2) and Fig.4 (b-3). The unusual events are marked in blue. These graphs show the following:

- Fig.4 (b-2) reveals that the pattern for relative humidity sensor 10 is different from the patterns for its neighborhood sensors (1, 26 and 143), which are all similar;
- Fig.4 (b-3) reveals that the pattern of air temperature sensor 10 is different from the patterns for its neighborhood sensors (1, 26 and 143), which are all similar.

These results indicate that the relative humidity and air temperature both changed simultaneously at sensor node 10. Therefore, we can assume that an unusual event happened at node 10.

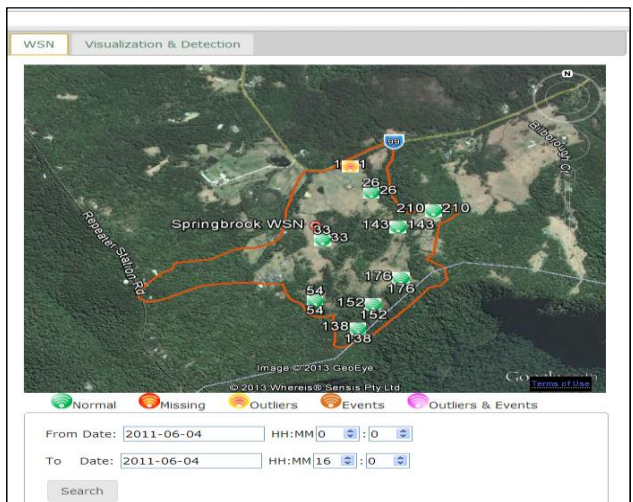

(a-1): Searching data and representing detection results in Google Earth

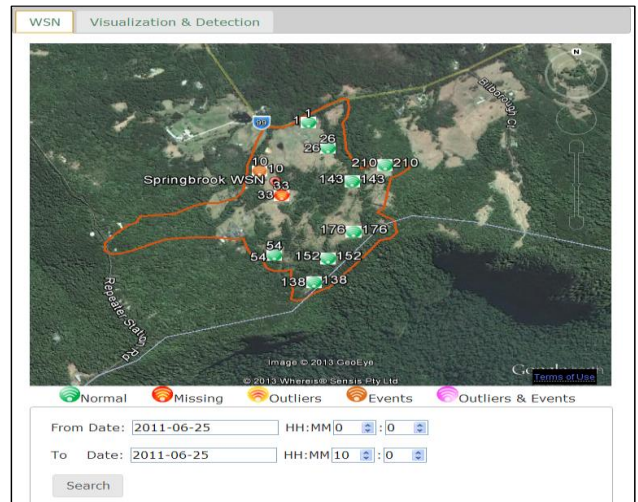

(b-1): Searching data and representing detection results in Google Earth

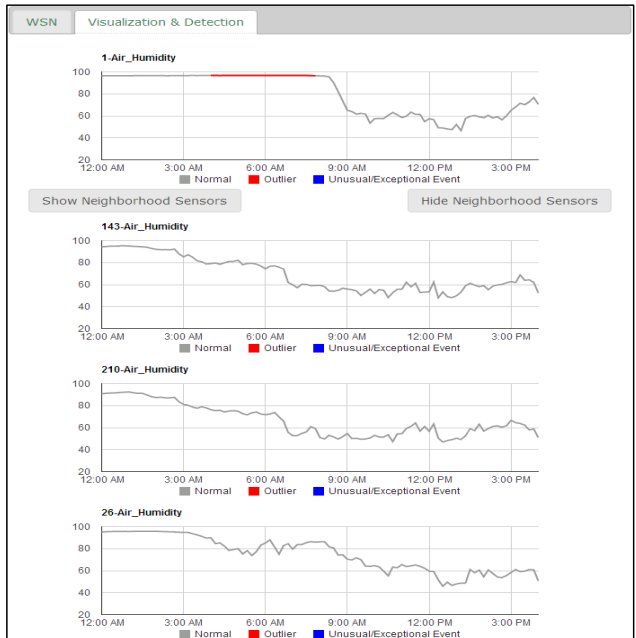

(a-2): Timeline user interface with segment outliers

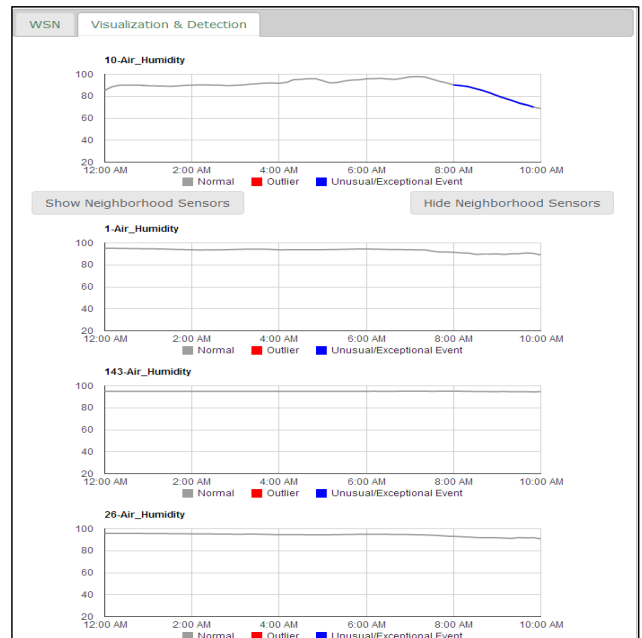

(b-2): Timeline user interface with unusual events

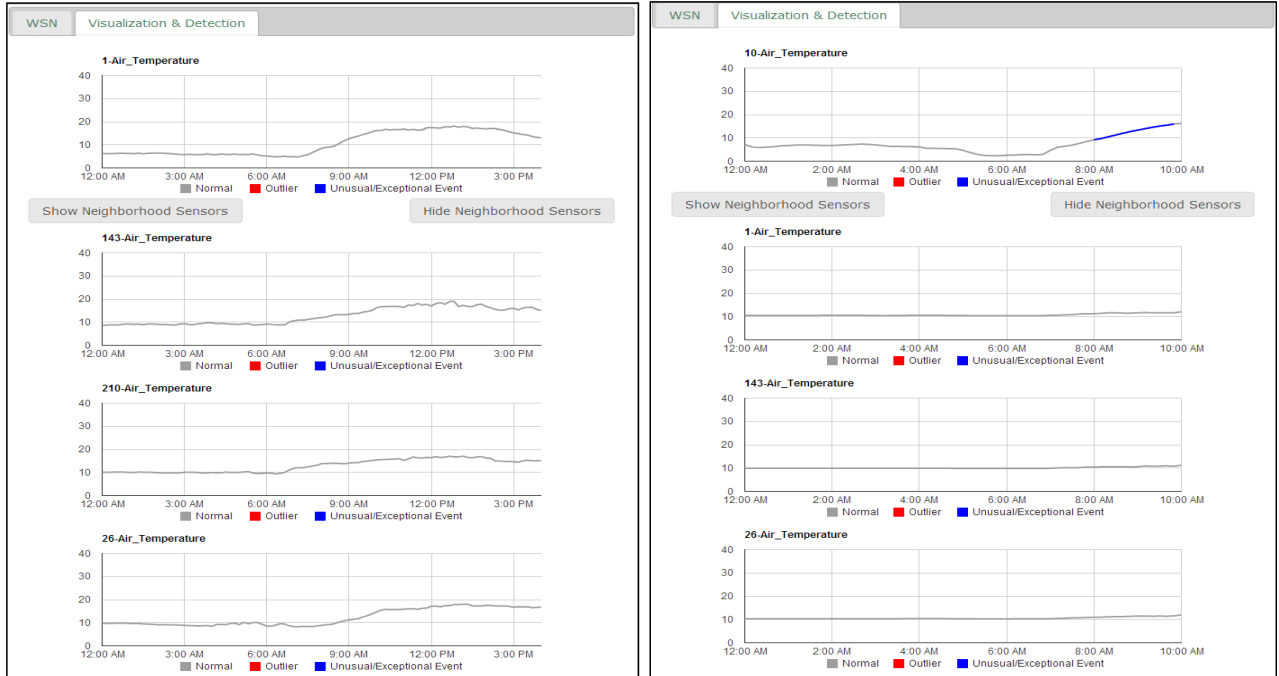

(a-3): Timeline user interface with normal data   (b-3): Timeline user interface with unusual events

Fig.4: Screenshots of user interfaces of SOUE-Detector

# 7. EVALUATION

## 7.1 Evaluation Metrics

To conduct our evaluations, we collected a real dataset from 36 sensor nodes deployed in the Springbrook project (Wark et al., 2008) as shown in Fig.5. For each sensor node, three different sensors are deployed - air temperature sensor, relative humidity sensor and air pressure sensor. The sampling rate is 1 sample/10 mins. Because we do not have ground truth information for segment outliers and unusual events for the real datasets, we generated three test datasets by adding segment outliers and unusual events to a cleaned real dataset that does not contain any abnormal data for the period 2011-08-18 - 2011-09-18.

The first test dataset is created by adding erroneous segment outliers to the cleaned dataset. The generation process comprises two steps:
1. Firstly, we randomly selected 8 air temperature data streams that do not overlap temporally. Next, for each selected data stream, we chose a segment with 12 observations and replaced this segment data stream with an air temperature segment outlier, generated by randomly choosing 12 values (between 0 and 25 C). The relative humidity and air pressure data streams are unchanged. Next, we repeated the first step 15 times across the duration of the segment (1 month).
2. Secondly, we randomly selected 8 air humidity data streams that do not overlap temporally. For each selected data stream, we chose a segment with 12 observations and replaced this segment with a relative humidity segment outlier generated by randomly choosing 12 values (between 40 and 100%). The air temperature and air pressure data streams are unchanged. Next, we repeated the second step 15 times across the duration of the segment (1 month).

The first test dataset contains 155,520 relative humidity observations, 155,520 air temperature observations, and 155,520 air pressure observations, and 720 segment outliers.

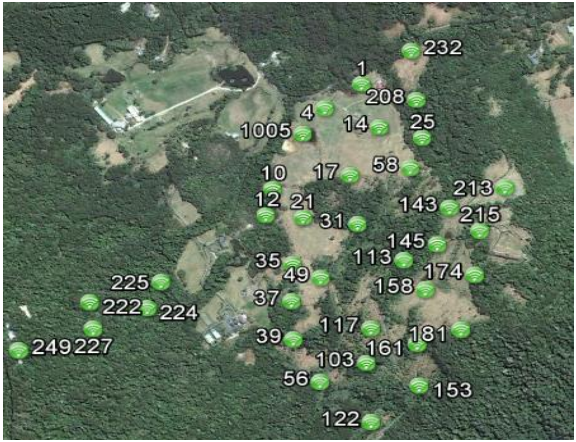

Fig.5: 36 sensor nodes deployed in the Springbrook National Park

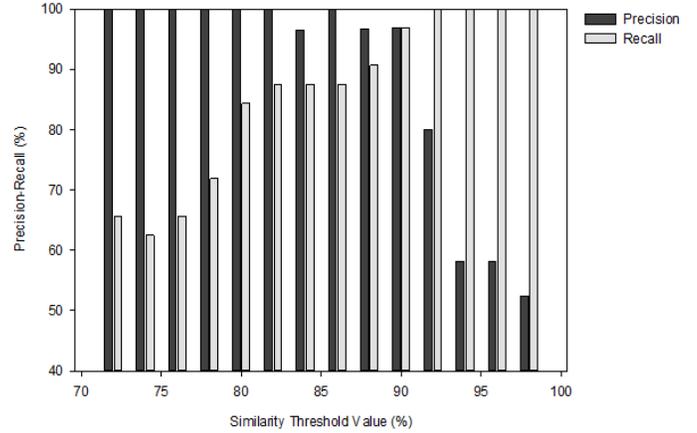

Fig. 6: Precision and Recall for detecting Erroneous Segment Outliers ($\partial = 300$ m, $\eta = 12$)

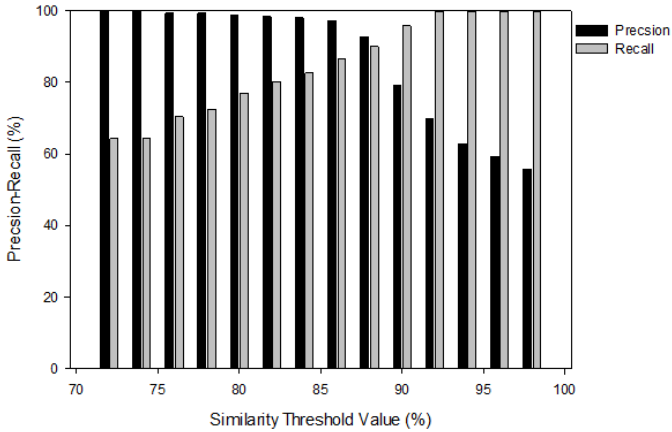

Fig. 7: Precision and Recall for detecting Unusual Events with *cesp:hasStrongCorrelation* ($\partial = 300$ m, $\eta = 12$)

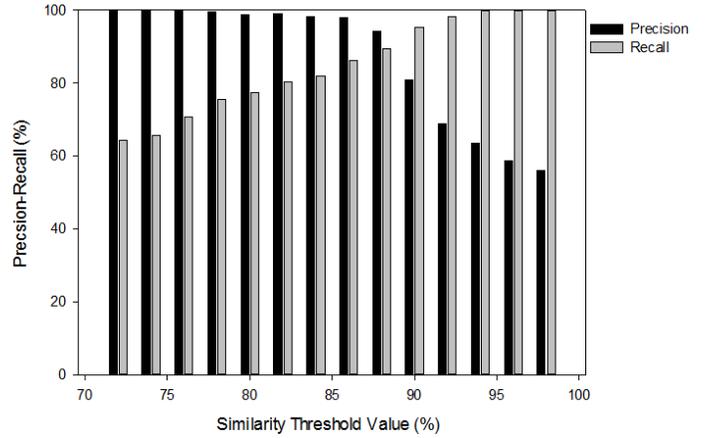

Fig. 8: Precision and Recall for detecting Unusual Events with *cesp:hasPositiveCorrelation* ($\partial = 300$ m, $\eta = 12$)

The remaining two test datasets are created by adding unusual events with two types of correlations, including *cesp:hasStrongCorrelation,* and *cesp:hasPositiveCorrelation,* to the cleaned dataset. Specifically, to create each test dataset, we randomly selected 8 sensor nodes. For each sensor node, we chose the relative humidity and air temperature data streams. Next, we chose a 2-hour time period and replaced the relative humidity and the air temperature segment data streams collected during this given 2-hour time period with 12 artificially generated relative humidity values and 12 artificially generated air temperature values. Moreover, the air pressure data streams are unchanged. Next, we repeated the above steps 30 times. Finally, each test dataset contains 155,520 relative humidity observations, 155,520 air temperature observations, and 155,520 air pressure observations, and 1,440 unusual events. In addition, each test dataset contains one type of unusual event.

### 7.2 Evaluation Results and Discussions

For our detection algorithm, we used a similarity threshold value to decide whether two segment sensor observations are similar or not. If the similarity value is greater than or equal to the similarity threshold value, they are regarded as similar. For both experiments, we set $\partial = 300$ m (predefined neighborhood threshold value) and $\eta = 12$ (predefined sliding window length). $\partial = 300$ guarantees that every sensor has at least 3 neighborhood sensors and a window of 12 samples has optimum potential to capture segment outliers and unusual events. Fig.6 shows the precision and recall for detecting erroneous segments from test data set 1 and Figs.7 to 8 respectively show the precision and recall for detecting unusual events from test data set 2 to test data set 3. The results from Fig.6 show that the recall is very high (87.59-100%) when using a similarity threshold greater than 81%. However, the precision of detecting segment outliers decreases when the similarity threshold increases above 92%. The results from Figs.7 to 8 reveal that the recall of detecting unusual events is very high (85.6-100%) when using a similarity threshold greater than 85%. However, the precision of detecting unusual events decreases as the similarity threshold rises above 90%.

In the evaluation, we evaluated the performance of both the outlier and unusual event detection methods using similarity threshold values β in the range of 70%-98%. The evaluation results show that the performance of our approach is very sensitive to the similarity threshold values. More specifically, the results indicate that setting the similarity threshold value at 90% gains the best performance for erroneous outlier detection, while setting the similarity threshold value at 88% gains the best performance for unusual event detection. More generally, in order to obtain optimum performance, similarity threshold values should lie in the range 85-92%. Overall, the evaluation results reveal that our approach is able to efficiently and accurately detect both erroneous outliers and unusual events by making use of sensor data trend similarities and correlations between sensor properties. In other words, such evaluation results reveal that integrating Semantic Web technologies and statistical algorithms with domain expert knowledge about sensor property correlations can improve the detection of outliers and unusual events within the sensor data wireless sensor data streams.

# 8. CONCLUSION AND FUTURE WORK

In this paper, we have developed a SOUE-Detector system that can efficiently detect genuine outliers and unusual events for real WSN data streams by combining DTW statistical analyses with domain expert knowledge and Semantic-Web based rules. In addition, the Web-based visualization interfaces enable scientists to quickly explore and easily understand the quality of their collected data streams. This can provide useful information for scientists in adjusting their WSN configurations to collect more accurate data streams. In addition, the Protégé user interface enables domain experts to record their knowledge about sensor property correlations that can be used to distinguish between erroneous outliers and events. Compared with previous related work, our approach has the following advantages. Firstly, we take into account multivariate sensor data and the relations between the variables. Secondly, our approach considers other types of correlation apart from just spatio-temporal correlations. Lastly, our approach handles the challenges associated with changes to wireless sensor network configurations over time.

To date, we have only evaluated our approach on test datasets that have been artificially generated and that contain outliers that are randomly generated. Future work aims include evaluating our algorithms and system on real data streams from the Springbrook and other WSNs that have not been cleaned and contain both erroneous data streams and unusual events. We also plan on linking the SOUEDetector with the Springbrook network "gateway" system to enable real time detection of segment outliers and unusual event detection and subsequent generation of notification services for decision-making.

# ACKNOWLEDGEMENTS
The work presented in this paper is supported by the China Scholarship Council. The authors would also like to thank Mr Luke Hovington and Mr Darren Moore from the CSIRO ICT Center for their explanations regarding sensor observations and relationships between sensor properties, as well as their valuable support and feedback.